\documentclass[pre,aps,reprint,twocolumn,footinbib,superscriptaddress,longbibliography,floatfix]{revtex4-1}
\usepackage{graphicx}
\usepackage{dcolumn}
\usepackage{bm}
\usepackage{amssymb}
\usepackage{pifont}
\usepackage{amsmath}
\usepackage{times}

\begin{document}

\title{
Volcano transition 
in populations of phase oscillators with random nonreciprocal interactions
}
\author{Diego Paz\'o}
\affiliation{Instituto de F\'{i}sica de Cantabria (IFCA), Universidad de
Cantabria-CSIC, 39005 Santander, Spain}
\author{Rafael Gallego}
\affiliation{Departamento de Matem\'aticas, Universidad de Oviedo, Campus de
Viesques, 33203 Gij\'on, Spain}
\date{\today}

 \begin{abstract}
Populations of heterogeneous phase oscillators with frustrated random
interactions exhibit a quasi-glassy state in which the distribution of local
fields is volcano-shaped. 
In a recent work [Phys.~Rev.~Lett.~{\bf 120}, 264102 (2018)] the volcano 
transition was replicated in a solvable model using a low-rank, random coupling matrix $\mathbf M$. 
We extend here that model including tunable nonreciprocal interactions, 
i.e.~${\mathbf M}^T\ne \mathbf M$. More specifically, we formulate two different solvable models.
In both of them the volcano transition persists if matrix elements $M_{jk}$ and $M_{kj}$ 
are enough correlated.
Our numerical simulations fully confirm the analytical results.  
To put our work in a wider context, we also investigate numerically the volcano transition
in the analogous model with a full-rank random coupling matrix.
\end{abstract}

\maketitle 

\section{introduction} Spin glasses are paradigmatic complex systems, whose
study found application in other seemingly unrelated fields, from optimization
problems to biology \cite{MPV87}. 
In 1992, Daido modified the Kuramoto model of phase oscillators replacing 
uniform ferromagnetic-like interactions
by random frustrated couplings \cite{Dai92}, 
exactly as in the Sherrington-Kirkpatrick spin-glass model \cite{SK75,MPV87}.  
The presence of frustrated interactions was expected to result in some sort 
of `oscillator glass'.
It was argued in \cite{Dai92} that the onset of a quasi-glassy phase
---characterized by algebraic relaxation and ``quasientrainment''---  
coincided with a reconfiguration of the local fields, such that their density adopted a volcano shape 
(with maximal density away from zero).  
This conclusion was the subject of some controversy  \cite{SR98,Dai00,SR00},
see also Sec.~IV.B.1.~of \cite{ABP+05}.
However, recent numerical simulations by Kimoto and Uezu  \cite{kimoto19} 
provided additional support to the  glassy nature of the volcano phase 
by measuring a suitably defined spin-glass order parameter.

In parallel to the previous works, phase oscillator ensembles 
endowed with low-rank random coupling matrices
have been studied \cite{bonilla93,KLS14,uezu15,OS18},
with the expectation that they reproduce features of the original full-rank-disordered
system \cite{Dai92}. In particular, Ottino-L\"offler and Strogatz \cite{OS18}
replicated the volcano transition with associative-memory-type interactions. 
In contrast to the original setup \cite{Dai92}, the model in \cite{OS18} 
does not display algebraic relaxation dynamics typical of spin glasses, 
but it has the advantage of being analytically solvable. 
A similar model had been {previously} analyzed 
with a completely different approach by Uezu and coworkers \cite{uezu15},
exploiting a theoretical link between the oscillator
population and the classical XY model. The results in \cite{uezu15}
and \cite{OS18} are complementary and mutually consistent;
still, the approach in \cite{OS18} has the advantage 
of making stability analysis possible.
 
In real spin glasses, as well as in {the} Daido model \cite{Dai92}, 
the interactions are symmetric, i.e.~reciprocal.  However, nonreciprocal interactions 
are found almost everywhere, from aggregates of neurons to self-motile active particles, see
e.g.~\cite{fruchart21}.  Populations of phase oscillators with asymmetric
couplings are found in models inspired in neuroscience \cite{MP18,laing21},
society \cite{HS11}, hydrodynamically coupled flagella \cite{uchida10},
etc. The effect of nonreciprocity on glassy phases in the context of synchronization
is attracting attention \cite{hanai22} but remains scarcely explored; particularly
in comparison to random neural networks, see e.g.~the discussion in \cite{dani18,mongillo} and references therein.
Remarkably, incorporating asymmetric couplings in {the} Daido model has only been undertaken by Stiller and Radons 
in \cite{SR98}. In this work it was concluded that the quasi-glassy phase did not persist 
if the random interactions were not reciprocal enough (in statistical sense). Still, 
it is important to stress that what is called quasi-glassy in \cite{SR98} differs from the state
emerging at the volcano transition in \cite{Dai92}. Revisiting these questions
appears to be in order, specially considering the current computational power.

In this paper we put forward two solvable models
of populations of oscillators with low-ranked, asymmetric, random interactions.
Specifically, {we} generalize the model in \cite{OS18} {by} introducing a free
parameter $\eta\in[-1,1]$, which allows us to continuously interpolate between
fully symmetric ($\eta=1$) and fully antisymmetric interactions ($\eta=-1$),
going over the uncorrelated case ($\eta=0$).  This new ingredient does not
degrade the tractability of the models. Moreover, we refine the analysis in
\cite{OS18} and allow the frequency distribution to be any
unimodal symmetric distribution (not only Lorentzian). 
Surprisingly, in spite of the similarities of the models introduced here, their phase
diagrams turn out to be notably different. 
For comparison purposes, we carry out simulations with the equivalent
model with a full-rank random coupling matrix \cite{SR98}. We find that 
the volcano transition is only possible above a critical level 
of reciprocity, different from the low-rank models. Perhaps the main message 
of this work is the impossibility of extrapolating the quantitative results 
from low-rank  to full-rank structural disorder, refuting a conjecture 
raised in \cite{OS18}.

This article is organized as follows. In Sec.~II we introduce the
two models of populations of phase oscillators 
with low-rank asymmetric coupling matrix.
Section III is devoted to the numerical study of the volcano transition 
in both models. The results are theoretically described
in Sec.~IV. Section V presents a numerical study of the volcano transition
for full-rank structural disorder with reciprocal and nonreciprocal 
interactions. Finally, Sec.~VI summarizes  
the main conclusions of this work. 

\section{Models with low-rank coupling matrix}

We investigate a population of heterogeneous phase oscillators
with quenched random couplings:
\begin{equation}
{\dot\theta}_j = \omega_j + \frac{J}N \sum_{k=1}^N M_{jk}
\sin(\theta_k-\theta_j).
\label{model}
\end{equation}
The phases $\theta_j$ are cyclic variables,  
and the population size is $N\gg1$. The natural frequencies $\{\omega_j\}_{j=1,\ldots,N}$ 
are drawn from a symmetric unimodal distribution $g(\omega)$, which is assumed
to be centered at zero without lack of generality (by going to a rotating frame if necessary).
In Eq.~\eqref{model}, matrix elements $M_{jk}$ codify the competition between 
synchronizing ($M_{jk}>0$) and
anti-synchronizing ($M_{jk}<0$) interactions.
Moreover, we include a global coupling constant $J>0$. 

In \cite{OS18} the coupling matrix ${\mathbf M}$ was constrained 
to be symmetric. Here we introduce a parameter $\eta$ 
controlling the weight in ${\mathbf M}$ of the symmetric and the antisymmetric matrices, ${\mathbf S}$ and ${\mathbf A}$, 
respectively. We have therefore:
\begin{equation}
{\mathbf M}=\frac12 \left[(1+\eta) {\mathbf S} + (1-\eta) {\mathbf A} \right] \, . 
\label{M}
\end{equation}
We allow parameter $\eta$ to vary in the range $[-1,1]$. 
The symmetric situation is recovered for $\eta=1$.

\subsection{Model 1}

The first model we propose assumes that 
each oscillator has two associated $L$-dimensional connectivity vectors 
${\bf u}_j$ and ${\bf v}_j$,
with quenched random components equal to $\pm1$: ${\bf u}_{j},{\bf v}_{j} \in \{\pm1\}^L$.
The elements of the symmetric ($\mathbf S$) and antisymmetric ($\mathbf A$) matrices are
computed from scalar products of the interaction vectors:
\begin{subequations}
\label{SA}
\begin{eqnarray}
S_{jk}&=& {\bf u}_j \cdot {\bf u}_k - {\bf v}_j \cdot {\bf v}_k \, ,  \label{S}\\
A_{jk}&=&{\bf u}_j \cdot {\bf v}_k - {\bf v}_j \cdot {\bf u}_k \, . \label{A}
\end{eqnarray}
\end{subequations}
Note that, with this formulation,  self-interactions 
are automatically excluded: $S_{jj}=A_{jj}=0$.
Matrix $\bf S$ yields the coupling type already used in \cite{OS18} (with a slightly different definition),
while $\bf A$  codifies an anti-reciprocal interaction.

The statistical properties of {offdiagonal} elements of the coupling matrix
are summarized next. 
First of all, the
mean is zero ($\langle M_{jk}\rangle=0$). The variance is
\begin{equation}
\langle M_{jk}^2 \rangle= L (1+\eta^2) \, .
\label{var}
\end{equation}
The correlation {coefficient between} mirror elements above and below the main diagonal of ${\mathbf M}$ is
\begin{equation}
 \mbox{corr}(M_{jk},M_{kj})= 
 \frac{\langle M_{jk} M_{kj} \rangle}{\langle M_{jk}^2 \rangle}
 =\frac{2\eta}{1+\eta^2} \, .
 \label{corr}
\end{equation}
The correlation vanishes at $\eta=0$, while
maximal (anti)correlation is achieved at $\eta=1$ ($-1$).

The rank of matrix $\mathbf M$ is $2L$, save for $\eta=0$ \footnote{in that case
$\mathrm{rank}({\mathbf M})=L$ since $M_{jk}=\frac12({\bf u}_j - {\bf
v}_j)\cdot ({\bf u}_k + {\bf v}_k)$.}. 
For the sake of analytical tractability, 
we assume low-ranked disorder, mathematically expressed
by the condition $L\ll\log_2N$. This key assumption permits us to
use the Ott-Antonsen ansatz \cite{OA08}, as in \cite{OS18}.

\subsection{Model 2}

Our second model is similar to model 1, but with independent interaction vectors for matrix $\mathbf A$.
In this way we have:
\begin{subequations}
\label{SAnew}
\begin{eqnarray}
S_{jk}&=& {\bf u}_j \cdot {\bf u}_k - {\bf v}_j \cdot {\bf v}_k \, ,  \\
A_{jk}&=&\tilde {\bf u}_j \cdot \tilde {\bf v}_k - \tilde {\bf v}_j \cdot \tilde {\bf u}_k \, . \label{Anew}
\end{eqnarray}
\end{subequations}
This means that each oscillator is characterized by its frequency $\omega_j$  and
four $L$-dimensional vectors (${\bf u}_j,{\bf v}_j,\tilde {\bf u}_j, \tilde {\bf v}_j$).
Variance and correlation in Eqs.~\eqref{var} and \eqref{corr} also hold for model 2. 
Now, however, matrices $\bf S$ and $\bf A$ are statistically independent.
The rank of matrix $M$ is $4L$, save for $\eta=\pm1$ where it equals $2L$.

\section{Numerical Results}
In our simulations we adopt a Gaussian probability density function
for the natural frequencies:
$$
g(\omega)=\frac{e^{-\frac{\omega^2}{2\sigma^2}}}{\sigma\sqrt{2\pi}}. 
$$
The value of the standard deviation $\sigma$
can be arbitrarily selected by rescaling time and the coupling constant $J$
in Eq.~\eqref{model}. We adopt $\sigma=\sqrt{\pi/2}$ for the numerical simulations.
Throughout this paper the numerical integration of the ordinary differential equations
is performed using the fourth-order Runge-Kutta method with time step 0.1 t.u.

\begin{figure}
  \includegraphics[width=0.9\columnwidth]{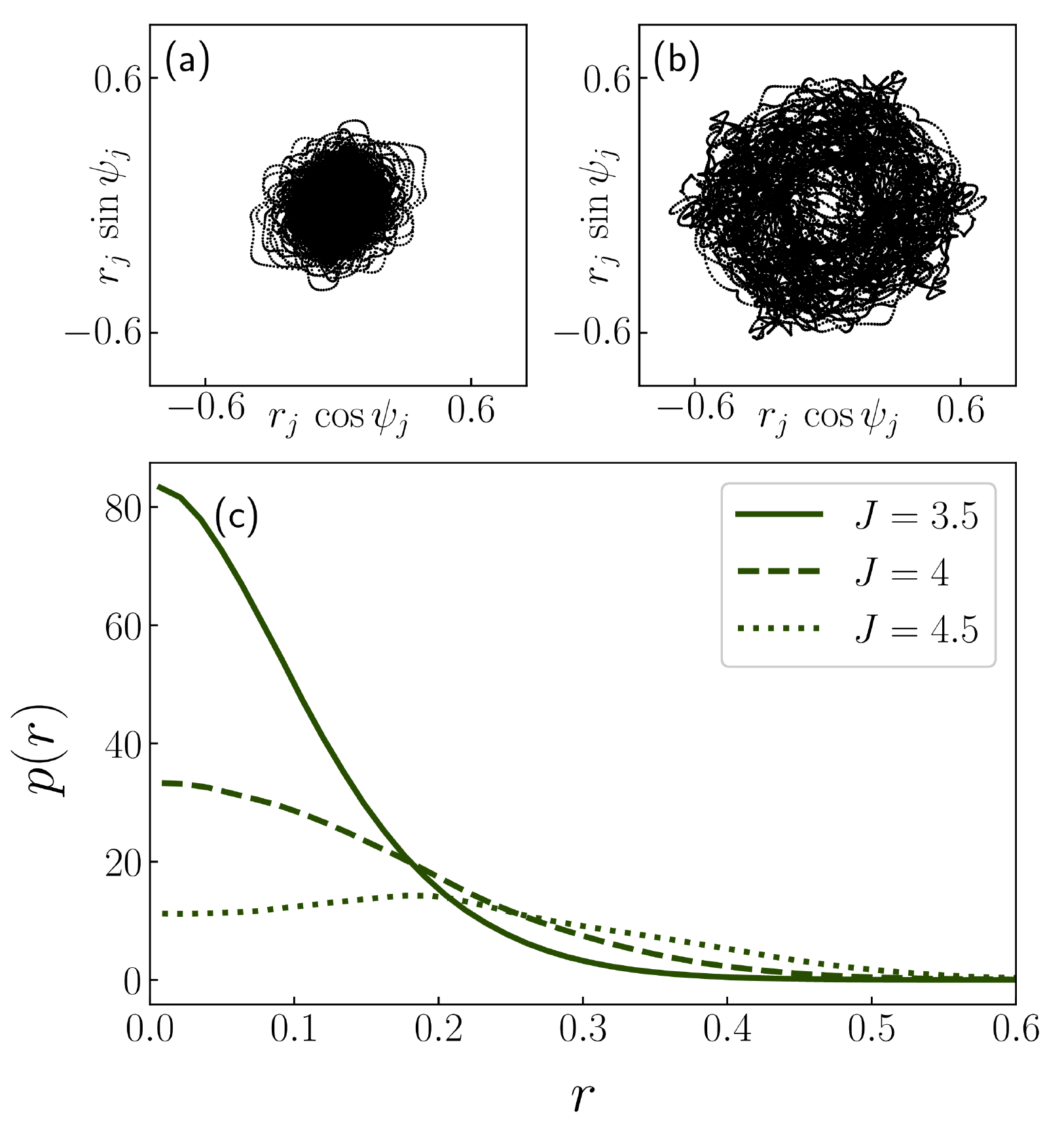}
  \caption{Phase portraits of the local fields $P_j(t)$ below and above the volcano transition:
  (a) $J=3.5$, and (b) $J=4.5$. In graph (c) we show the radial distributions of local
  fields for the previous values of $J$ together with an intermediate
  value ($J=4$).}
  \label{fig::D}
\end{figure}

\subsection{Model 1}
We start considering model 1.
For fixed $\sigma$, our system depends on four free parameters:
$N$ (the population size), $L$ (the dimension of the interaction vectors 
$\mathbf u$ and $\mathbf v$), 
$J$ (the coupling constant), and $\eta$ (the asymmetry parameter).
We are interested in the large-$N$ behavior, such that only a marginal 
dependence upon the realization of frequencies and connections
is expected. 
The effect of the remaining parameters $L$, $J$ and $\eta$ is 
investigated hereafter.

Similar to the spin-glass transition, the emergence of the volcano phase
cannot be detected measuring a global order parameter. It remains
near zero below and above the critical point. Instead, it is the 
distribution of complex local fields
\begin{equation}
P_j(t)\equiv r_j e^{i\psi_j}= \frac1N\sum_{k=1}^N M_{jk} e^{i\theta_k} \, ,
\label{lf}
\end{equation}
what undergoes a structural change, see below.
For later use it is convenient to rewrite Eq.~\eqref{model} in terms of the local fields:
\begin{equation}
{\dot\theta}_j = \omega_j + J r_j \sin(\psi_j-\theta_j) \, .
\label{vom}
\end{equation}

Our first numerical simulation, in Figs. 1(a) and 1(b),
show the phase portraits of the local fields $P_j(t)$
for a population of $N=500$ oscillators interacting 
nonreciprocally with asymmetry parameter $\eta=1/4$ and $L=3$.
Figures~1(a) and 1(b) correspond to values of $J$
below and above the volcano transition, respectively. 
In the former plot the density of local fields is
maximal at the origin, while in Fig.~1(b) the
volcano shape is apparent.
Figure 1(c) depicts the radial distributions of local fields
for the same values of $J$ as in Figs.~1(a) and 1(b), plus an intermediate
$J$ value near the critical value $J_v$.
At the volcano transition the radial distribution of the local fields
$p(r)$ changes from concave down at the origin to concave up. This means
that $p(r)$ peaks at $r_*>0$ for $J>J_v$.

\begin{figure}
  \includegraphics[width=0.9\columnwidth]{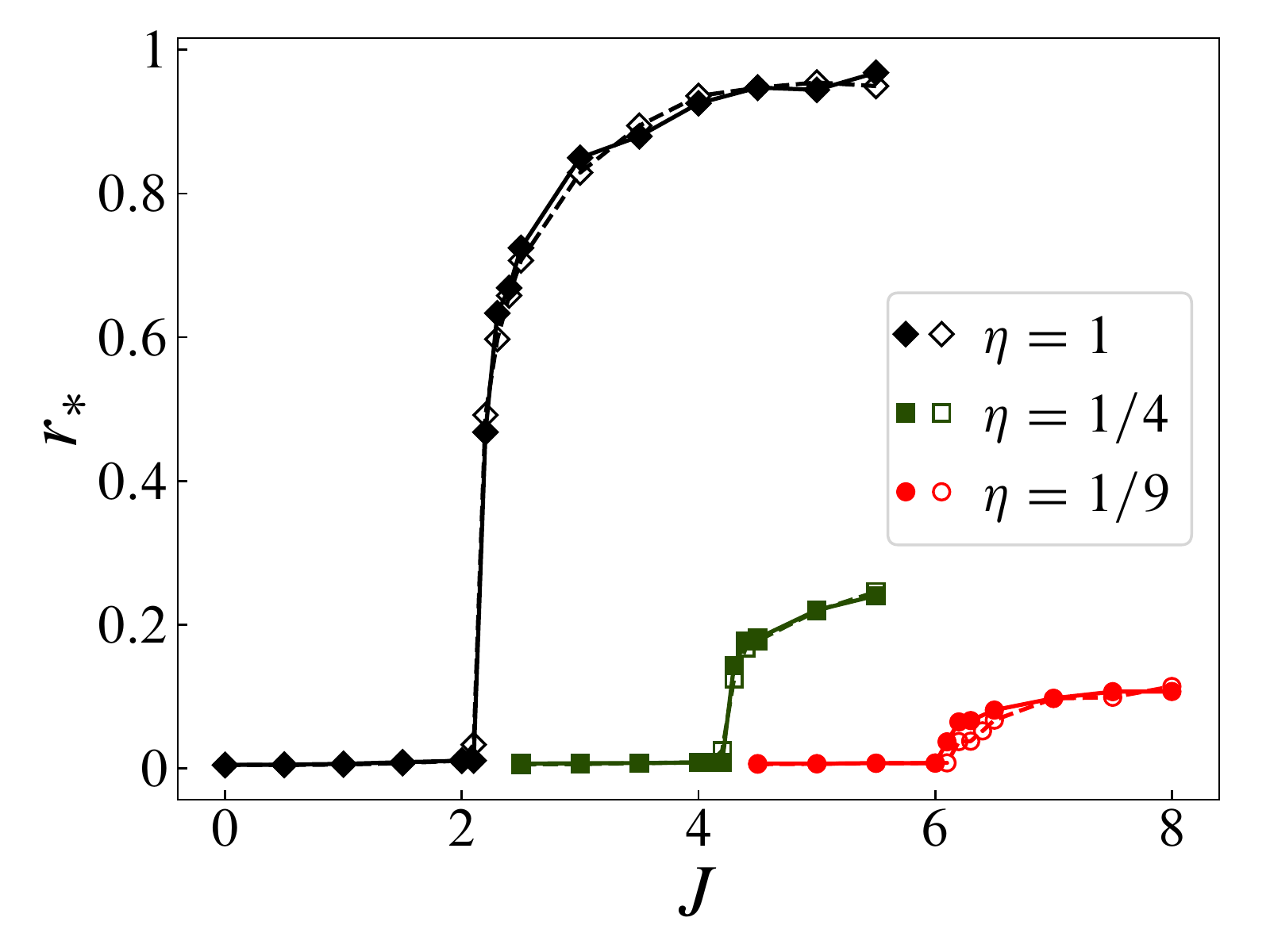}
  \caption{Position of the maximum in the radial distribution of local fields as a function
  of the coupling constant $J$ for model 1 with $N=500$ oscillators.
Data sets correspond to three different values of $\eta\in\{1, 1/4,1/9\}$. For each $\eta$ 
value two different connectivity-vector sizes were selected:
$L=2$ (filled symbols connected by a solid line) and
$L=3$ (empty symbols, dashed line). Each point 
represents the location of the maximum in the histogram of the local-field amplitudes,
collected from 100 realizations of random frequencies and connectivity vectors.
For each simulation the system was integrated for $500$ t.u., after a transient of $300$ t.u.}
  \label{fig::PT}
\end{figure}

Next, we investigate the dependence of $r_*$ on the coupling constant $J$. 
Three values of $\eta\in\{1,1/4,1/9\}$ were selected.
In addition, for each of them two different values of the vector dimension $L\in\{2,3\}$
were chosen. Figure \ref{fig::PT} presents the results. 
The peak radius $r_*$ departs from zero above the $\eta$-dependent critical coupling $J_v(\eta)$. Notably, $J_v$ increases as $\eta$ is lowered,
i.e.~as the correlation between mirror entries of the connectivity matrix
decreases. Moreover, $r_*$ attains significantly smaller values as $\eta$
is lowered, i.e.~the ``volcano width'' decreases as the interactions become less reciprocal. 
Eventually, no transition is found for $\eta\le0$. This means that the volcano transition
requires positive correlations between $M_{jk}$ and $M_{kj}$ to occur.  It is
also interesting to note that the results are apparently insensitive to the size
$L$ of the connectivity vectors, as already pointed out in
\cite{OS18} for $\eta=1$. Actually, the irrelevance of $L$ only holds provided that
$2^{2L}\ll N$, as theoretically justified in Sec.~IV.

\subsection{Model 2}

\begin{figure}
  \includegraphics[width=0.9\columnwidth]{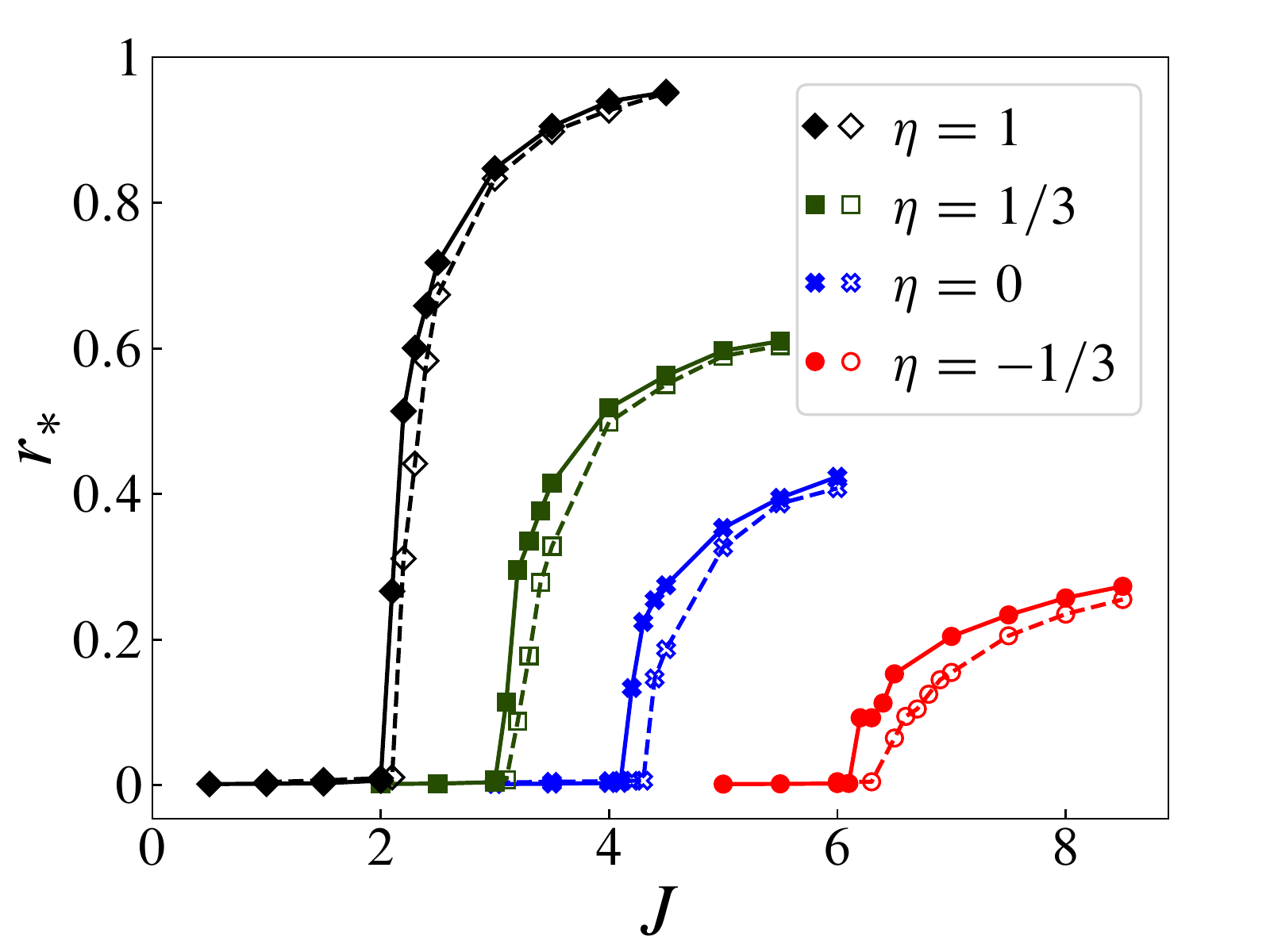}
  \caption{Position of the maximum in the radial distribution of local fields as a function
  of the coupling constant $J$ for model 2 with $N=1000$ oscillators.
Data sets correspond to four different values of $\eta\in\{-1/3,0,1/3,1\}$. For each $\eta$ 
value simulations adopted two different connectivity-vector sizes:
$L=1$ (filled symbols connected by solid lines)
and $L=4$ (empty symbols, dashed lines).}
  \label{fig::PT2}
\end{figure}

Our numerical study of model 2 proceeded analogously to model 1. 
As above, we tracked 
the peak value $r_*$ as a function of the coupling constant $J$, 
for different values of $\eta$ and $L$. 
The results for $N=1000$ are shown in Fig.~\ref{fig::PT2}. 
As with model 1, the smaller $\eta$ the larger the critical coupling $J_v$.
To our surprise, the volcano transition also occurs for negative $\eta$ values,
i.e.~when the inwards and the outwards connections are statistically anticorrelated.
When $L$ is changed from $1$ to $4$, a small displacement of the values of $J_v(\eta)$ is observed in the figure.
We argue in the next section that $J_v$ is insensitive to the value of $L$, 
provided that $2^{4L}\ll N$, This does not hold at all for $L=4$ ($2^{16}\approx6.5 \times 10^4$), 
but nevertheless the displacement of $J_v$ in this case remains quite moderate.

\section{Theoretical Analysis}

The theoretical analysis is similar for models 1 and 2. 
In both cases we adopt the thermodynamic limit $N\to \infty$,
such that the state of the system is described 
by a phase density $\rho$.

\subsection{Model 1}
We start presenting the theory for model 1, and 
defer the relevant modifications for model 2 to the end of this section.
To lighten the notation we introduce a $2L$-dimensional vector $\bf w$,
which is the concatenation of the interaction vectors 
$\bf u$ and $\bf v$: ${\bf w}=({\bf u}^T,{\bf v}^T)^T$.
Adopting this notation, we have that 
$\rho(\theta|\omega,\mathbf{w};t) d\theta$ is the fraction 
of oscillators with phases between $\theta$ and $\theta+d\theta$ 
at time $t$ with natural frequency $\omega$ and interaction vector $\bf w$.

The density obeys the continuity
equation:
\begin{equation}
\partial_t \rho+ \partial_\theta (v \rho)=0 .
\label{cont}
\end{equation}
Here, recalling Eq.~\eqref{vom}, we use $v$ as a short-hand notation for the
velocity:
\begin{equation}
v(\theta,\omega,{\bf w},t)=\omega+ J \, \mathrm{Im} [P({\bf w},t) e^{-i\theta}] \, . 
\label{v}
\end{equation}
The local field $P$, see Eq.~\eqref{lf},
is simply the double average of $e^{i\theta}$ over $2^{2L}$
different interaction vectors and over the continuum of natural frequencies:
\begin{multline}
P({\bf w},t)= \\ \frac{1}{2^{2L}} \sum_{\bf w '}
{\cal{M}}({\bf w},{\bf w}') 
\int_{-\infty}^\infty d\omega\,  g(\omega)   \int_0^{2\pi} \rho(\theta|\omega,{\bf w}';t) e^{i\theta} d\theta .\label{P0}
\end{multline}
Matrix $\bm{\mathcal{M}}$ has dimension $2^{2L}\times 2^{2L}$, and its elements are calculated as 
those of ${\mathbf M}$, see Eq.~\eqref{M}.
Each row (column) corresponds to a 
different binary string $\bf w$ (${\bf w}'$), hence the matrix dimension.
{(We define $2^{2L}\times 2^{2L}$ matrices ${\bm{\mathcal{S}}}$ and ${\bm{\mathcal{A}}}$
from Eq.~\eqref{SA} analogously.)}
The dependence of $v$ on $\rho$ via $P$ confers nonlinearity 
to the continuity Eq.~\eqref{cont}.

The key point of the analysis it the fact that, as Eq.~\eqref{v} only depends on the first harmonic in $\theta$,
we can apply the Ott-Antonsen ansatz \cite{OA08}:
\begin{equation}
 \rho(\theta|\omega,{\bf w}; t)= \frac{1}{2\pi}\left[1+\sum_{n=1}^\infty  \alpha(\omega,{\bf w},t)^n e^{i n\theta} + \mbox{c.c.}\right] \,.
 \label{oa}
\end{equation}
Here, $\alpha$ is the coefficient of the first harmonic, and c.c.~stands for complex conjugate.
Inserting the previous expansion into the continuity equation \eqref{cont}, we obtain 
the evolution equation for $\alpha$:
\begin{equation}
\partial_t \alpha(\omega,{\bf w},t)=-i\omega\alpha+\frac{J}2 \left[P^* -P \alpha^2 \right].
\label{alpha}
\end{equation}
Moreover, assuming the Ott-Antonsen ansatz \eqref{oa}, the equation
for the local field~\eqref{P0} simplifies:
\begin{equation}
P({\bf w},t)=\frac{1}{2^{2L}}\sum_{{\bf w} '}{\cal M}({\bf w},{\bf w}') \int
d\omega\,  g(\omega) 
\alpha(\omega,{\bf w }',t)^* .
\label{P}
\end{equation}
Plugging this expression into Eq.~\eqref{alpha}, we get a closed 
vector integro-differential equation.

We analyze next the stability of the incoherent state $\alpha=0$
against infinitesimal perturbations.
Therefore we drop the nonlinear term in Eq.~\eqref{alpha}.
For the resulting linear system, we take an exponential ansatz 
$\alpha(\omega,{\bf w},t)=\beta(\omega,{\bf w}) e^{\lambda t}$.
This yields an equation for the exponential growth rate $\lambda$:
\begin{equation}
(\lambda +i\omega) \beta(\omega,{\bf w})= 
\frac{J}{2^{2L+1}}\sum_{{\bf w} '}{\cal M}({\bf w},{\bf w}') b({\bf w}'),
\end{equation}
where we have introduced the shorthand notation $b({\bf w})\equiv\int d\omega  g(\omega) \beta(\omega,{\bf w }) $.
Reordering terms, and integrating over $\omega$ both sides of the equation, we get:
\begin{equation}
 b({\bf w})=\frac{J}{2^{2L+1}} \int_{-\infty}^{\infty} \frac{g(\omega)}{\lambda+i\omega} d\omega 
 \sum_{{\bf w} '}{\cal M}({\bf w},{\bf w}')  b({\bf w}')
\label{sc}
 \end{equation}
At the critical coupling $J_v$, the real part of the eigenvalue $\lambda=\lambda_r+i\Omega$
approaches zero: $\lambda_r\to0^+$. Hence, the previous equation at
criticality becomes (written in matrix form):
\begin{equation}
\left(c \frac{J_v}{2^{2L+1}} {\bm{\mathcal{M}}}  - {\bm{\mathcal{I}}} \right) {\mathbf b}=0,
\label{cjmib}
\end{equation}
where
$c\equiv\pi g(-\Omega) - i \int g(\omega-\Omega)/\omega\, d\omega$,
and ${\bm{\mathcal{I}}}$ is the identity matrix.
Nontrivial solutions (${\mathbf b}\ne{\mathbf 0}$) of the linear equation \eqref{cjmib}
are eigenvectors of $\bm{\mathcal{M}}$ corresponding to nonzero eigenvalues. 

Computing the eigenvalues of 
$\bm{\mathcal{M}}$ is surprisingly simple.
We find that ${\bm{\mathcal{M}}}^2=\eta {\bm{\mathcal{S}}}^2$,
by virtue of the identities
${\bm{\mathcal{S}}}^2+{\bm{\mathcal{A}}}^2={\bm{\mathcal{SA}}}+{\bm{\mathcal{AS}}}=0$, which are easily proven. 
Therefore, all we need is 
the eigenvalue spectrum of ${\bm{\mathcal{S}}}$. As found in \cite{OS18},
the nonzero eigenvalues are 
$2^{2L}$ and $-2^{2L}$, both of them with multiplicity
$L$.
Hence, 
the nonzero eigenvalues of ${\bm{\mathcal{M}}}$ 
are simply $\pm\sqrt{\eta}\, 2^{2L}$. 
For $\eta<0$ the nontrivial eigenvalues are purely 
imaginary, and no solution of Eq.~$\eqref{cjmib}$ exists since $\mathrm{Re}(c)\ne0$, 
i.e.~there is not a critical $J_v$ value.
For $\eta>0$, matrix ${\bm{\mathcal{M}}}$ possesses real eigenvalues
and Eq.~\eqref{cjmib} may only hold provided $\mathrm{Im}(c)=0$. If $g(\omega)$ is an even unimodal function, then 
$\Omega=0$ necessarily.
In turn, $c=\pi g(0)$, and the critical coupling 
turns out to be:
\begin{equation}
J_v^{(1)}=\frac{1}{\sqrt\eta} \times \frac{2}{\pi g(0)}. 
\label{jc}
\end{equation}
We have included the superscript $(1)$ to emphasize the result refers to model 1.
Equation \eqref{jc} should provide an accurate estimation
of $J_v$ for large ensembles of oscillators, if the condition 
$N\gg2^{2L}$ is fulfilled. This condition is tantamount to
assuming that a large number of 
oscillators share each of the possible interaction vectors ${\bf w}$, 
such that the continuous formulation is meaningful.
Let us confront Eq.~\eqref{jc} with the result of the numerical simulations 
condensed in Fig.~\ref{fig::PT}. 
For a frequency dispersion $\sigma=\sqrt{\pi/2}$, Eq.~\eqref{jc} becomes $J_v=2/\sqrt{\eta}$.
In particular, for the values $\eta$ selected in Fig.~2, 
the predicted critical couplings are $J_v=2$, $4$, and $6$, irrespective of $L$. 
We observe an excellent agreement between theory and simulations. 

\begin{figure}
  \includegraphics[width=0.9\columnwidth]{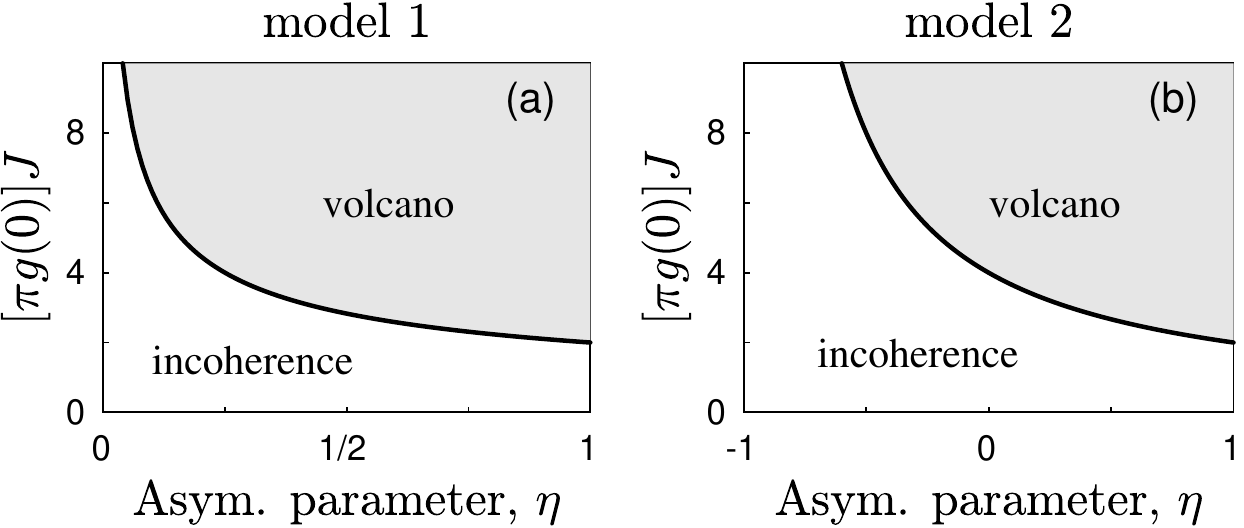}
  \caption{Phase diagrams of models 1 (a) and 2 (b) for symmetric unimodal frequency distribution.
  The boundaries of the volcano phase in panels (a) and (b)
  are defined by Eqs.~\eqref{jc} and \eqref{jc2}, respectively. Note the different range of $\eta$ in each panel.}
  \label{fig::PD}
\end{figure}

We end the analysis of model 1 showing its phase diagram 
in Fig.~\ref{fig::PD}(a). Notice the divergence of the volcano phase boundary
as $\eta\to0^+$, i.e.~as the correlation between inward and outward links vanishes.

\subsection{Model 2}

The theoretical analysis of model 2 is analogous to the one for model 1 above. Therefore, we only
indicate the key differences.
We have now four $L$-dimensional vectors associated with each oscillator. 
In turn, the dimensionality of matrix $\bm{\mathcal{M}}$ is $2^{4L}\times2^{4L}$, 
as there are $2^{4L}$ different combinations of $\bf u$, $\bf v$, $\tilde{\bf u}$ and $\tilde{\bf v}$.
The mathematical relation between $\bm{\mathcal{M}}$ and the $2^{2L}\times2^{2L}$ matrices $\bm{\mathcal{S}}$
and $\bm{\mathcal{A}}$ is not trivial at first sight. It can be conveniently expressed
with the Kronecker product, denoted by $\otimes$:
\begin{equation}
\bm{\mathcal{M}}=\frac{1}{2} \left[ (1+\eta)\bm{\mathcal{E}}\otimes \bm{\mathcal{S}}+ (1-\eta) \bm{\mathcal{A}}\otimes \bm{\mathcal{E}}   \right],
\label{m2}
\end{equation}
where $\bm{\mathcal{E}}$ is a $2^{2L}\times2^{2L}$ matrix of ones.

Eventually, the analysis leads to a marginality condition analogous to Eq.~\eqref{cjmib}:
\begin{equation}
\left(c \frac{J_v}{2^{4L+1}} {\bm{\mathcal{M}}}  - {\bm{\mathcal{I}}} \right) {\mathbf b}=0 \,,
\label{cjmib_new}
\end{equation}
where $ {\bm{\mathcal{I}}}$ is the $2^{4L}\times2^{4L}$ identity matrix. 
The critical coupling is 
dictated by the eigenvalue of ${\bm{\mathcal{M}}}$ with the largest real part.
Therefore, the problem reduces to 
finding the eigenvalue spectrum of ${\bm{\mathcal{M}}}$.

Firstly, let us note that  $\bm{\mathcal{E}}\otimes\bm{\mathcal{S}}$ commutes with $\bm{\mathcal{A}}\otimes\bm{\mathcal{E}}$,
implying they share a common basis of eigenvectors.
 The proof follows: 
 $({\bm{\mathcal{E}}}\otimes{\bm{\mathcal{S}}})
(\bm{\mathcal{A}}\otimes\bm{\mathcal{E}})-
(\bm{\mathcal{A}}\otimes\bm{\mathcal{E}})
(\bm{\mathcal{E}}\otimes\bm{\mathcal{S}})=
\bm{\mathcal{E A}}\otimes\bm{\mathcal{S E}}-\bm{\mathcal{AE}}\otimes\bm{\mathcal{E S}}=0$,
where we have used that $\bm{\mathcal{E}}$ 
commutes with $\bm{\mathcal{S}}$ and $\bm{\mathcal{A}}$ in the last identity.

At this point we notice that ${\bm{\mathcal{E}}}$ possesses only one nonzero eigenvalue $2^{2L}$, 
with associated eigenvector ${\bf e}_1=(1,1, \ldots, 1)^T$. Hence, the relevant 
eigenvectors of ${\bm{\mathcal{M}}}$ have the form 
${\mathbf e}_1 \otimes \bm {\mathbf s}_i$ or
${\mathbf a}_i \otimes \bm {\mathbf e}_1$, where 
$\{{\mathbf s}_i\}$ and $\{{\mathbf a}_i\}$ are the eigenvector sets corresponding to nonzero eigenvalues
of $\bm{\mathcal{S}}$ and $\bm{\mathcal{A}}$, 
respectively. Other combinations of eigenvectors yield null eigenvalues; 
note in particular that $\bm{\mathcal{E}} {\mathbf s}_i=0$,
because the ${\mathbf s}_i$'s are orthogonal to ${\mathbf e}_1$. Likewise, $\bm{\mathcal{E}} {\mathbf a}_i=0$.

Following our previous discussion, the eigenvalue
spectrum of ${\bm{\mathcal{M}}}$ is easily obtained.  Nonzero 
eigenvalues come from either term of $\bm{\mathcal{M}}$, see
Eq.~\eqref{m2}, by virtue of the identities $\bm{\mathcal{S}}
\mathbf{e}_1=\bm{\mathcal{A}} \mathbf{e}_1=\mathbf 0$.  (Matrices
$\bm{\mathcal{S}}$ and $\bm{\mathcal{A}}$ have zero row sum.) Eigenvalues
corresponding to the eigenvectors ${\mathbf a}_i \otimes \bm {\mathbf e}_1$ are
pure imaginary, since $\bm{\mathcal{A}}$ is a
skew-symmetric matrix. Their exact values {are}
therefore immaterial for this problem \footnote{For completeness: The imaginary eigenvalues
of ${\bm{\mathcal{M}}}$ are $\pm i (1-\eta) 2^{4L-1}$. A sketch of the 
proof follows. We denote the normalized eigenvectors of $\bm{\mathcal{S}}$
with positive (negative) eigenvalue by ${\mathbf s}_i^+$ (${\mathbf s}_i^-$) .
From the identities ${\bm{\mathcal{S}}}^2+{\bm{\mathcal{A}}}^2={\bm{\mathcal{SA}}}+{\bm{\mathcal{AS}}}=0$,
we find that ${\bm{\mathcal{A}}} {\mathbf s}_i^{+}=2^{2L} {\mathbf s}_i^{-}$ 
and ${\bm{\mathcal{A}}} {\mathbf s}_i^{-}=- 2^{2L} {\mathbf s}_i^{+}$.
The eigenvectors of ${\bm{\mathcal{A}}}$ are of the form ${\mathbf s}_i^{+}\pm i{\mathbf s}_i^{-}$,
and the associated eigenvalues are $\pm i 2^{2L}$, completing in this way the proof.}.  The other nonzero
eigenvalues of ${\bm{\mathcal{M}}}$, corresponding to eigenvectors ${\mathbf e}_1 \otimes \bm {\mathbf s}_i$,
are real. Their values are $\pm(1+\eta) 2^{4L-1}$, since we know 
$\bm{\mathcal{S}} {\mathbf s}_i= {\pm} 2^{2L} {\mathbf s}_i$ from \cite{OS18}. The critical coupling is obtained by considering the
positive eigenvalue of $\bm{\mathcal{M}}$ in Eq.~\eqref{cjmib_new}:
\begin{equation}
J_v^{(2)}=\frac{4}{1+\eta} \times \frac{1}{\pi g(0)} 
\label{jc2}
\end{equation}
This equation accurately predicts the volcano transition in our simulations
in Fig.~\ref{fig::PT2}: For the four values of $\eta$ selected ($1$, $1/3$, $0$, and $-1/3$),
Eq.~\eqref{jc2} predicts $J_v=2$, $3$, $4$, and $6$.

Equation \eqref{jc2} allows us to represent the phase diagram in Fig.~\ref{fig::PD}(b).
Remarkably, the divergence of $J_v$ is now located at $\eta_\infty=-1$, contrasting
with $\eta_\infty=0$ in model 1.
This discrepancy implies that the correlation between $M_{jk}$ and $M_{kj}$ 
is not enough to determine the value of $\eta_\infty$. The key difference between models 1 and
2 is that  
the symmetric and antisymmetric matrices $\mathbf S$ and $\mathbf A$ 
---contributing to $\mathbf M$---
are independent for model 2, but not for model 1.

 \subsection{Model 1+2}

We conclude the theoretical analysis noticing that models 1 and 2 can be combined to create 
a continuum of solvable models, with two antisymmetric components
weighted by parameter $\beta$:
\begin{equation}
A_{jk}=(1-\beta)[{\bf u}_j \cdot {\bf v}_k - {\bf v}_j \cdot {\bf u}_k]
+\beta [\tilde{\bf u}_j \cdot \tilde{\bf v}_k - \tilde{\bf v}_j \cdot \tilde{\bf u}_k] 
\label{beta} 
\end{equation}
This model remains analytically tractable with critical coupling 
\begin{equation}
J_v=\frac{4}{\sqrt{4\eta+(2\beta-\beta^2)(1-\eta)^2}} \times \frac{1}{\pi g(0)}  \, . 
\end{equation}
The divergence of $J_v$ occurs at $\eta_\infty=\beta/(\beta-2)$.

\section{Model with full-rank coupling matrix}
\label{sec::fr}
In this section we investigate to what extent our previous results carry over to the 
original model with a full-rank coupling matrix \cite{Dai92,SR98,kimoto19}.
Such a possibility was already explored in \cite{OS18} for reciprocal coupling
and Lorentzian $g(\omega)$ with an inconclusive answer.

The ``full-rank model'' writes:
\begin{equation}
{\dot\theta}_j = \omega_j + \frac{\tilde J}{\sqrt{N}} \sum_{k=1}^N J_{jk}
\sin(\theta_k-\theta_j) \, .
\label{modeld}
\end{equation}
Nondiagonal elements of the coupling matrix $\mathbf J$
are drawn from a zero-mean, unit-variance Gaussian distribution.
Note the  prefactor $N^{-1/2}$,
instead of $N^{-1}$ as in \eqref{model}.
The correlation between symmetric elements ($j\ne k$) is:
\begin{equation}
\langle J_{jk} J_{kj}\rangle=  \tau \, .
\end{equation}

As above, we focus on the radial density of the 
local fields, which are 
$$
r_j e^{i\Psi_j}=\frac{1}{\sqrt{N}} \sum_k J_{jk} e^{i\theta_k} \,.
$$

\subsection{Reciprocal coupling, $\tau=1$}

We start with the symmetric case ($\tau=1$), originally considered
in \cite{Dai92}, and recently revisited in \cite{kimoto19}. 
In the latter work an analytical value of $\tilde J_v$ 
was proposed based on a mapping between model \eqref{modeld} with
Gaussian $g(\omega)$ and the classical XY model at finite temperature.

The thermodynamic limit of model \eqref{modeld} 
coincides with our model in Eq.~\eqref{model}
adopting $L=N/2\to\infty$, since the binomial-distributed matrix elements become
Gaussian distributed (central limit theorem). 
As done in \cite{OS18}, it is worth extrapolating our 
analytical result in Eq.~\eqref{jc}, setting $\eta=1$, to 
the full-rank model \eqref{modeld}, even if it is out of its range of validity.
The extrapolated critical coupling turns out to be
quite simple:
\begin{equation}
\tilde J_v= \frac{2}{\pi g(0)} \, .
\label{jct}
\end{equation}
Remarkably, if we particularize this result for Gaussian $g(\omega)$ 
the value of $\tilde J_v$ coincides with the one theorized in Ref.~\cite{kimoto19}.
The latter reference exploited a mapping between the saddle-point equations of the classical XY model
at finite temperature and the model in Eq.~\eqref{modeld} with Gaussian $g(\omega)$. 
Still, the numerical verification of the result in \cite{kimoto19}
was not straightforward. The numerical procedure involved 
decreasing the frequency dispersion quasi-adiabatically. 

\begin{figure}
  \includegraphics[width=0.9\columnwidth]{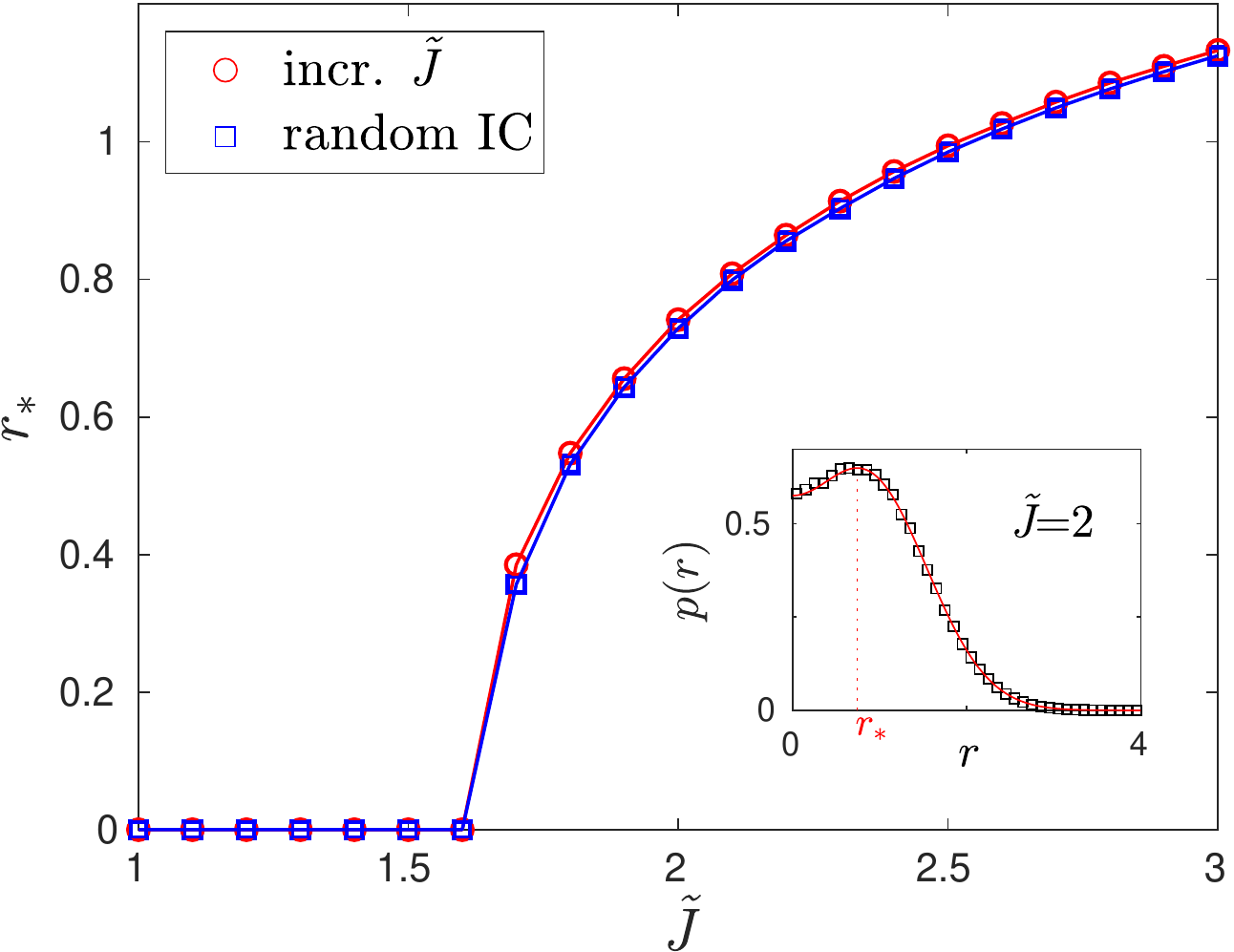}
  \caption{Volcano transition in the model defined by Eq.~\eqref{modeld} for $\tau=1$ and $N=400$. 
Red circles and blue squares indicate the position of the maximum in the radial distribution of local fields  
$r_*$, for two different protocols: increasing $\tilde J$ quasi-adiabatically (see text), and with 
random initial phases for each $\tilde J$ value. 
The exact value of $r_*$ for each $\tilde J$ value is determined fitting the histogram values.
Inset: Empirical radial distribution of local fields for $\tilde J=2$. The red curve is the 
biparametric fitting function in Eq.~\eqref{fit}.}
  \label{fig::PTF}
\end{figure}

To shed more light on this issue, we decided to implement the simulation ourselves. 
Instead of slowly decreasing $\sigma$ as in \cite{kimoto19}, 
we kept $\sigma=\sqrt{\pi/2}$ slowly increasing $\tilde J$ from 0.
Both procedures should be almost equivalent if the increments in $\tilde J$
are small enough
and transients are long enough.
In our case, we implemented steps of size $\Delta \tilde J=10^{-3}$, 
and integrations $10^3$ t.u.~long for each $\tilde J$ value. At particular values of $\tilde J$ the system was integrated for 
$5\times 10^{3}$ t.u.~saving the local fields every time unit. This procedure was 
repeated 20 times, each with an independent sampling of natural frequencies and coupling matrix elements
in order to achieve good statistics. 
The behavior of the peak value $r_*$ for $N=400$ is shown in Fig.~\ref{fig::PTF}
(virtually the same result is obtained for $N=200$). 
In the figure 
the red circles are the maxima of the fitting function used to smooth
the histogram. 
For each $\tilde J$ value, the histogram was fitted 
by the normalized two-parametric 
function 
\begin{equation}
p(r)=C \, \{\exp[-(r-\mu)^2/\xi]+\exp[-(r+\mu)^2/\xi]\} \, , 
\label{fit}
\end{equation}
where 
$C(\mu,\xi)$ is the normalization constant: $\int_0^\infty p(r) r dr=1$.
The form of $p(r)$ is suggested by the solution for $\tilde J=0$: $\mu=0$,
$\xi=1$, and Ref.~\cite{OS18}. The goodness of the fittings is excellent for all $\tilde J$ values.
This is illustrated, for $\tilde J=2$, by the inset in Fig.~\ref{fig::PTF}.
From the fitting line, the condition $p'(r_*)=0$, allows to obtain the maximum
as the nontrivial solution of $e^{4\mu r_*/\xi} (\mu-r_*)-\mu-r_*=0$.
The results in Fig.~\ref{fig::PTF} show a critical point
clearly below the value $\tilde J=2$, predicted by Eq.~\eqref{jct}.
Thus our result aligns with the work by Daido \cite{Dai92}, under appropriate
rescalings, but not with Ref.~\cite{kimoto19}.

The discrepancy with the result in \cite{kimoto19} is intriguing.
The theoretical treatment in \cite{kimoto19} is not completely 
rigorous, specially concerning the drifting oscillators, whose contribution 
is neglected.  {Regarding} the numerical procedure in \cite{kimoto19}, the results are obtained from one realization of the model, not from an ensemble of realizations.
{As a final effort to improve our understanding, we decided to redo the simulation in \cite{kimoto19}.
The frequencies were sampled once and their dispersion was progressively decreased, keeping the coupling matrix
elements $J_{jk}$ and the coupling constant ($\tilde J=1$) fixed.
Our result, for one single realization as in \cite{kimoto19}, is a critical dispersion $\sigma_v$ closer to 
our inference from the empirical value of ${\tilde J}_v$ in Fig.~5 ($\sigma_v=\sqrt{\pi/2}/{\tilde J}_v$) than 
to the prediction from Eq.~\eqref{jct} ($\sigma_v=\sqrt{\pi/8}$), see the Supplmental Material \footnote{See the Supplemental Material at [...] for details}.}
Whatever the 
{correct interpretation of these results is}, it is clear that the model with full-rank disorder
is truly more complex than the models with low-rank disorder.

\subsection{Nonreciprocal coupling, $\tau<1$}

Next, we consider the model defined by Eq.~\eqref{modeld} with nonreciprocal interactions, i.e.~$\tau<1$.
In \cite{SR98} this model was investigated, putting the focus on 
a static phase ---at large coupling--- in which all the oscillators are frozen in random positions 
(in a certain rotating frame).
Such state was identified with the spin glass (in disagreement with \cite{Dai92,kimoto19}).
We focus here on the volcano transition, without investigating the dynamics further.
We anticipate that the static phase found in \cite{SR98}
falls inside the volcano phase, see below.

\begin{figure}
  \includegraphics[width=0.9\columnwidth]{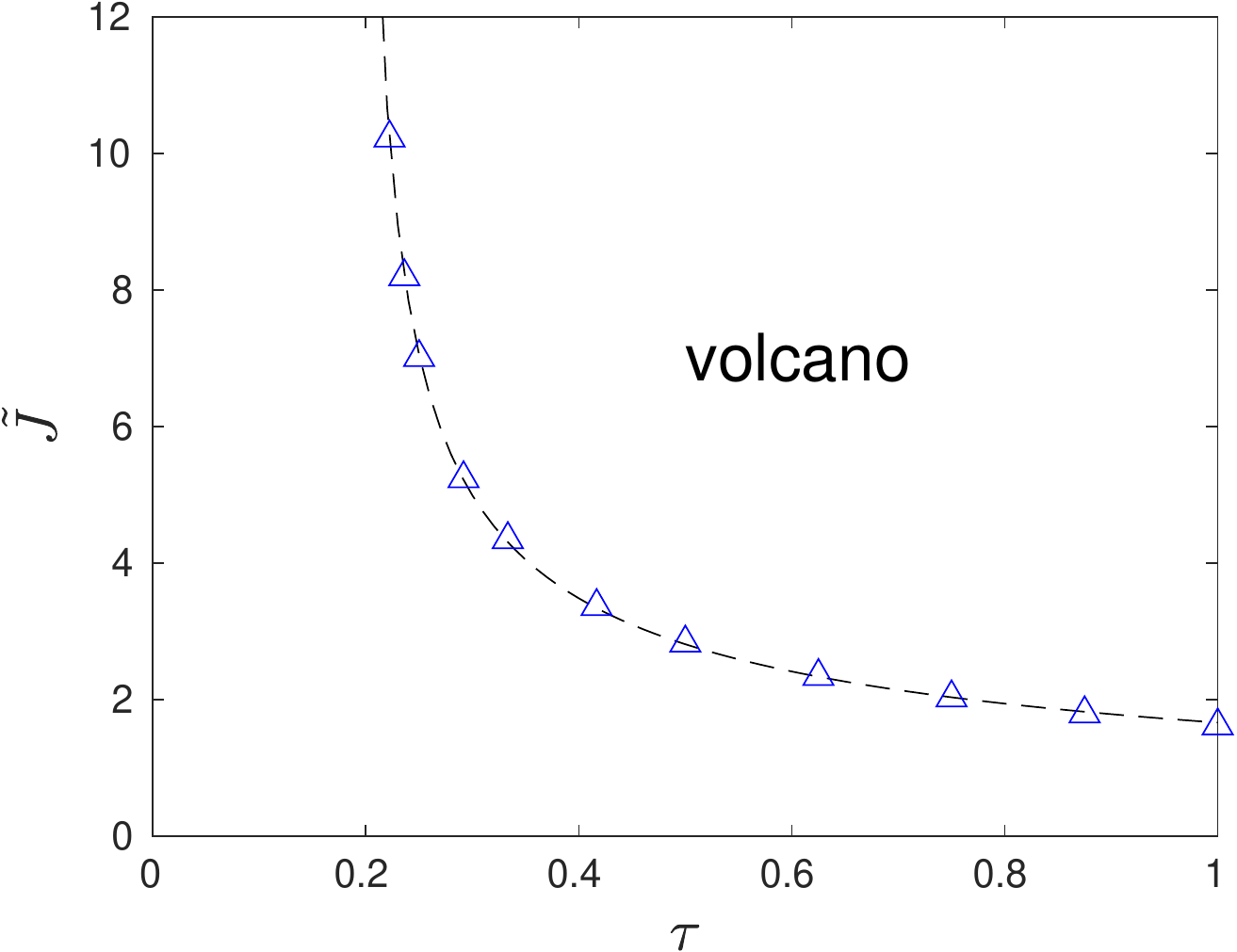}
  \caption{Empirical phase diagram of model \eqref{modeld} with $N=400$ oscillators, and 
  Gaussian frequency distribution of variance $\pi/2$. Triangles correspond
  to empirical critical couplings of the volcano transition $\tilde J_v$, at particular values of the correlation $\tau$.
  To determine $\tilde J_v$, the values of $\tilde J$ were sampled with a spacing of 0.1.
  The amplitudes of the local fields were collected from 100 independent realizations,
  each run for 1000 t.u.~after a transient of 2000 t.u.~and random initial phases.
  The resulting histograms of local fields radii were fitted to Eq.~\eqref{fit}.
  As a guide to the eye we depict a dashed line, obtained fitting 
  the critical couplings to the function $\tilde J_v(\tau)=a/(\tau-\tau_\infty)^\gamma$;
  {$\tau_\infty=0.1938$, $\gamma=0.542$, $a=1.481$.}}
  \label{fig::pd_full}
\end{figure}

Our initial numerical simulations for $\tau=1/4$
did not reveal any difference between increasing $J$ 
quasiadiabatically or setting random initial conditions (not shown),
as occurred with $\tau=1$. 
Hence, we decided to estimate the critical coupling $\tilde J_v(\tau)$ 
taking random initial conditions, irrespective of the value of $\tau$.
Figure~\ref{fig::pd_full} summarizes our results. Triangles
mark the location of the volcano transition at different $\tau$ values. 
The data are fitted to a simple algebraic formula,
suggesting a divergence of $\tilde J_v$ at at a critical 
$\tau$ value around 0.19, see figure caption.

Our two models with low-rank disorder and the model with full-rank disorder
have in common that the critical coupling increases as reciprocity is decreased. 
At the same time, we observed that the ``volcano width'' decreases 
as reciprocity diminishes in all cases.
However, the divergence of the critical coupling 
occurs at a different level of nonreciprocity in each model.
There is a remarkable lack of uniformity in this respect.

\section{Conclusions}

We conclude recapitulating the main findings in this work:

\begin{enumerate}
 \item The volcano transition is observed in populations of phase oscillators with 
nonreciprocal coupling. This applies to the two solvable models introduced here, as well
to the model defined by Eq.~\eqref{modeld}.

\item 
Nonreciprocity hinders the volcano transition. It may become even impossible,
but the critical level of reciprocity depends on the specific model.

\item 
Concerning reciprocal interactions, the results of our simulations with full-rank coupling
matrix do not agree with those in \cite{kimoto19}. We detect the volcano transition
at a critical coupling neatly below the one proposed in \cite{kimoto19}, 
which turns out to be exactly the value extrapolating the low-rank model to full-rank (i.e.~$L=N/2$).

\item Models with low-rank disorder may serve as a surmise for the full-rank case.
Still, the possibility of extrapolating from them, as speculated in \cite{OS18}, has proven to be 
overly optimistic. More sophisticated techniques, e.g.~based on the cavity method \cite{MPV87,mongillo}, 
await to be developed for phase oscillator ensembles.

\item Our work provides two different families of
nonsymmetric low-rank random matrices
with known eigenvalue spectra. This may be useful in other domains, 
such as recurrent neural networks in computational neuroscience \cite{mastrogiuseppe,schuessler}.

\end{enumerate}

We have focused on the volcano transition in systems of phase oscillators with
random connectivity, but more work is required to shed light on their dynamics,
not only the distribution of local fields.
In particular, the relationship between the volcano phase and the glassy dynamics deserves
further study. Our work also evidences that many interesting questions remain 
to be solved analytically. Looking back on past achievements, we remain moderately optimistic, 
even if nonreciprocity represents an additional difficulty.

\section*{Acknowledgments}

We acknowledge support by Grant No.~PID2021-125543NB-I00, funded by MCIN/AEI/10.13039/501100011033 and by ERDF A way of making Europe


\begin{thebibliography}{26}%
\makeatletter
\providecommand \@ifxundefined [1]{%
 \@ifx{#1\undefined}
}%
\providecommand \@ifnum [1]{%
 \ifnum #1\expandafter \@firstoftwo
 \else \expandafter \@secondoftwo
 \fi
}%
\providecommand \@ifx [1]{%
 \ifx #1\expandafter \@firstoftwo
 \else \expandafter \@secondoftwo
 \fi
}%
\providecommand \natexlab [1]{#1}%
\providecommand \enquote  [1]{``#1''}%
\providecommand \bibnamefont  [1]{#1}%
\providecommand \bibfnamefont [1]{#1}%
\providecommand \citenamefont [1]{#1}%
\providecommand \href@noop [0]{\@secondoftwo}%
\providecommand \href [0]{\begingroup \@sanitize@url \@href}%
\providecommand \@href[1]{\@@startlink{#1}\@@href}%
\providecommand \@@href[1]{\endgroup#1\@@endlink}%
\providecommand \@sanitize@url [0]{\catcode `\\12\catcode `\$12\catcode
  `\&12\catcode `\#12\catcode `\^12\catcode `\_12\catcode `\%12\relax}%
\providecommand \@@startlink[1]{}%
\providecommand \@@endlink[0]{}%
\providecommand \url  [0]{\begingroup\@sanitize@url \@url }%
\providecommand \@url [1]{\endgroup\@href {#1}{\urlprefix }}%
\providecommand \urlprefix  [0]{URL }%
\providecommand \Eprint [0]{\href }%
\providecommand \doibase [0]{http://dx.doi.org/}%
\providecommand \selectlanguage [0]{\@gobble}%
\providecommand \bibinfo  [0]{\@secondoftwo}%
\providecommand \bibfield  [0]{\@secondoftwo}%
\providecommand \translation [1]{[#1]}%
\providecommand \BibitemOpen [0]{}%
\providecommand \bibitemStop [0]{}%
\providecommand \bibitemNoStop [0]{.\EOS\space}%
\providecommand \EOS [0]{\spacefactor3000\relax}%
\providecommand \BibitemShut  [1]{\csname bibitem#1\endcsname}%
\let\auto@bib@innerbib\@empty
\bibitem [{\citenamefont {M{\'e}zard}\ \emph {et~al.}(1987)\citenamefont
  {M{\'e}zard}, \citenamefont {Parisi},\ and\ \citenamefont
  {Virasoro}}]{MPV87}%
  \BibitemOpen
  \bibfield  {author} {\bibinfo {author} {\bibfnamefont {M.}~\bibnamefont
  {M{\'e}zard}}, \bibinfo {author} {\bibfnamefont {G.}~\bibnamefont {Parisi}},
  \ and\ \bibinfo {author} {\bibfnamefont {M.~A.}\ \bibnamefont {Virasoro}},\
  }\href@noop {} {\emph {\bibinfo {title} {Spin glass theory and beyond: An
  Introduction to the Replica Method and Its Applications}}},\ \bibinfo
  {series} {World Scientific Lecture Notes in Physics}, Vol.~\bibinfo {volume}
  {9}\ (\bibinfo  {publisher} {World Scientific Publishing Company},\ \bibinfo
  {address} {Singapore},\ \bibinfo {year} {1987})\BibitemShut {NoStop}%
\bibitem [{\citenamefont {Daido}(1992)}]{Dai92}%
  \BibitemOpen
  \bibfield  {author} {\bibinfo {author} {\bibfnamefont {H.}~\bibnamefont
  {Daido}},\ }\bibfield  {title} {\enquote {\bibinfo {title} {Quasientrainment
  and slow relaxation in a population of oscillators with random and frustrated
  interactions},}\ }\href@noop {} {\bibfield  {journal} {\bibinfo  {journal}
  {Phys. Rev. Lett.}\ }\textbf {\bibinfo {volume} {68}},\ \bibinfo {pages}
  {1073--1076} (\bibinfo {year} {1992})}\BibitemShut {NoStop}%
\bibitem [{\citenamefont {Sherrington}\ and\ \citenamefont
  {Kirkpatrick}(1975)}]{SK75}%
  \BibitemOpen
  \bibfield  {author} {\bibinfo {author} {\bibfnamefont {D.}~\bibnamefont
  {Sherrington}}\ and\ \bibinfo {author} {\bibfnamefont {S.}~\bibnamefont
  {Kirkpatrick}},\ }\bibfield  {title} {\enquote {\bibinfo {title} {Solvable
  model of a spin-glass},}\ }\href {\doibase 10.1103/PhysRevLett.35.1792}
  {\bibfield  {journal} {\bibinfo  {journal} {Phys. Rev. Lett.}\ }\textbf
  {\bibinfo {volume} {35}},\ \bibinfo {pages} {1792--1796} (\bibinfo {year}
  {1975})}\BibitemShut {NoStop}%
\bibitem [{\citenamefont {Stiller}\ and\ \citenamefont {Radons}(1998)}]{SR98}%
  \BibitemOpen
  \bibfield  {author} {\bibinfo {author} {\bibfnamefont {J.~C.}\ \bibnamefont
  {Stiller}}\ and\ \bibinfo {author} {\bibfnamefont {G.}~\bibnamefont
  {Radons}},\ }\bibfield  {title} {\enquote {\bibinfo {title} {Dynamics of
  nonlinear oscillators with random interactions},}\ }\href {\doibase
  10.1103/PhysRevE.58.1789} {\bibfield  {journal} {\bibinfo  {journal} {Phys.
  Rev. E}\ }\textbf {\bibinfo {volume} {58}},\ \bibinfo {pages} {1789--1799}
  (\bibinfo {year} {1998})}\BibitemShut {NoStop}%
\bibitem [{\citenamefont {Daido}(2000)}]{Dai00}%
  \BibitemOpen
  \bibfield  {author} {\bibinfo {author} {\bibfnamefont {H.}~\bibnamefont
  {Daido}},\ }\bibfield  {title} {\enquote {\bibinfo {title} {Algebraic
  relaxation of an order parameter in randomly coupled limit-cycle
  oscillators},}\ }\href {\doibase 10.1103/PhysRevE.61.2145} {\bibfield
  {journal} {\bibinfo  {journal} {Phys. Rev. E}\ }\textbf {\bibinfo {volume}
  {61}},\ \bibinfo {pages} {2145--2147} (\bibinfo {year} {2000})}\BibitemShut
  {NoStop}%
\bibitem [{\citenamefont {Stiller}\ and\ \citenamefont {Radons}(2000)}]{SR00}%
  \BibitemOpen
  \bibfield  {author} {\bibinfo {author} {\bibfnamefont {J.~C.}\ \bibnamefont
  {Stiller}}\ and\ \bibinfo {author} {\bibfnamefont {G.}~\bibnamefont
  {Radons}},\ }\bibfield  {title} {\enquote {\bibinfo {title} {Self-averaging
  of an order parameter in randomly coupled limit-cycle oscillators},}\ }\href
  {\doibase 10.1103/PhysRevE.61.2148} {\bibfield  {journal} {\bibinfo
  {journal} {Phys. Rev. E}\ }\textbf {\bibinfo {volume} {61}},\ \bibinfo
  {pages} {2148--2149} (\bibinfo {year} {2000})}\BibitemShut {NoStop}%
\bibitem [{\citenamefont {Acebr\'on}\ \emph {et~al.}(2005)\citenamefont
  {Acebr\'on}, \citenamefont {Bonilla}, \citenamefont {P\'erez-Vicente},
  \citenamefont {Ritort},\ and\ \citenamefont {Spigler}}]{ABP+05}%
  \BibitemOpen
  \bibfield  {author} {\bibinfo {author} {\bibfnamefont {J.~A.}\ \bibnamefont
  {Acebr\'on}}, \bibinfo {author} {\bibfnamefont {L.~L.}\ \bibnamefont
  {Bonilla}}, \bibinfo {author} {\bibfnamefont {C.~J.}\ \bibnamefont
  {P\'erez-Vicente}}, \bibinfo {author} {\bibfnamefont {F.}~\bibnamefont
  {Ritort}}, \ and\ \bibinfo {author} {\bibfnamefont {R.}~\bibnamefont
  {Spigler}},\ }\bibfield  {title} {\enquote {\bibinfo {title} {The {K}uramoto
  model: A simple paradigm for synchronization phenomena},}\ }\href@noop {}
  {\bibfield  {journal} {\bibinfo  {journal} {Rev. Mod. Phys.}\ }\textbf
  {\bibinfo {volume} {77}},\ \bibinfo {pages} {137--185} (\bibinfo {year}
  {2005})}\BibitemShut {NoStop}%
\bibitem [{\citenamefont {Kimoto}\ and\ \citenamefont {Uezu}(2019)}]{kimoto19}%
  \BibitemOpen
  \bibfield  {author} {\bibinfo {author} {\bibfnamefont {T.}~\bibnamefont
  {Kimoto}}\ and\ \bibinfo {author} {\bibfnamefont {T.}~\bibnamefont {Uezu}},\
  }\bibfield  {title} {\enquote {\bibinfo {title} {Correspondence between phase
  oscillator network and classical {X}{Y} model with the same random and
  frustrated interactions},}\ }\href {\doibase 10.1103/PhysRevE.100.022213}
  {\bibfield  {journal} {\bibinfo  {journal} {Phys. Rev. E}\ }\textbf {\bibinfo
  {volume} {100}},\ \bibinfo {pages} {022213} (\bibinfo {year}
  {2019})}\BibitemShut {NoStop}%
\bibitem [{\citenamefont {Bonilla}\ \emph {et~al.}(1993)\citenamefont
  {Bonilla}, \citenamefont {{P{\'e}rez Vicente}},\ and\ \citenamefont
  {Rub{\'i}}}]{bonilla93}%
  \BibitemOpen
  \bibfield  {author} {\bibinfo {author} {\bibfnamefont {L.L.}\ \bibnamefont
  {Bonilla}}, \bibinfo {author} {\bibfnamefont {C.J.}\ \bibnamefont {{P{\'e}rez
  Vicente}}}, \ and\ \bibinfo {author} {\bibfnamefont {J.M.}\ \bibnamefont
  {Rub{\'i}}},\ }\bibfield  {title} {\enquote {\bibinfo {title} {Glassy
  synchronization in a population of coupled oscillators},}\ }\href@noop {}
  {\bibfield  {journal} {\bibinfo  {journal} {J. Stat. Phys.}\ }\textbf
  {\bibinfo {volume} {70}},\ \bibinfo {pages} {921--937} (\bibinfo {year}
  {1993})}\BibitemShut {NoStop}%
\bibitem [{\citenamefont {Kloumann}\ \emph {et~al.}(2014)\citenamefont
  {Kloumann}, \citenamefont {Lizarraga},\ and\ \citenamefont
  {Strogatz}}]{KLS14}%
  \BibitemOpen
  \bibfield  {author} {\bibinfo {author} {\bibfnamefont {I.~M.}\ \bibnamefont
  {Kloumann}}, \bibinfo {author} {\bibfnamefont {I.~M.}\ \bibnamefont
  {Lizarraga}}, \ and\ \bibinfo {author} {\bibfnamefont {S.~H.}\ \bibnamefont
  {Strogatz}},\ }\bibfield  {title} {\enquote {\bibinfo {title} {Phase diagram
  for the {K}uramoto model with van {H}emmen interactions},}\ }\href {\doibase
  10.1103/PhysRevE.89.012904} {\bibfield  {journal} {\bibinfo  {journal} {Phys.
  Rev. E}\ }\textbf {\bibinfo {volume} {89}},\ \bibinfo {pages} {012904}
  (\bibinfo {year} {2014})}\BibitemShut {NoStop}%
\bibitem [{\citenamefont {Uezu}\ \emph {et~al.}(2015)\citenamefont {Uezu},
  \citenamefont {Kimoto}, \citenamefont {Kiyokawa},\ and\ \citenamefont
  {Okada}}]{uezu15}%
  \BibitemOpen
  \bibfield  {author} {\bibinfo {author} {\bibfnamefont {T.}~\bibnamefont
  {Uezu}}, \bibinfo {author} {\bibfnamefont {T.}~\bibnamefont {Kimoto}},
  \bibinfo {author} {\bibfnamefont {S.}~\bibnamefont {Kiyokawa}}, \ and\
  \bibinfo {author} {\bibfnamefont {M.}~\bibnamefont {Okada}},\ }\bibfield
  {title} {\enquote {\bibinfo {title} {Correspondence between phase oscillator
  network and classical {XY} model with the same infinite-range interaction in
  statics},}\ }\href {\doibase 10.7566/JPSJ.84.033001} {\bibfield  {journal}
  {\bibinfo  {journal} {J. Phys. Soc. Jpn.}\ }\textbf {\bibinfo {volume}
  {84}},\ \bibinfo {pages} {033001} (\bibinfo {year} {2015})}\BibitemShut
  {NoStop}%
\bibitem [{\citenamefont {Ottino-L\"offler}\ and\ \citenamefont
  {Strogatz}(2018)}]{OS18}%
  \BibitemOpen
  \bibfield  {author} {\bibinfo {author} {\bibfnamefont {B.}~\bibnamefont
  {Ottino-L\"offler}}\ and\ \bibinfo {author} {\bibfnamefont {S.~H.}\
  \bibnamefont {Strogatz}},\ }\bibfield  {title} {\enquote {\bibinfo {title}
  {Volcano transition in a solvable model of frustrated oscillators},}\ }\href
  {\doibase 10.1103/PhysRevLett.120.264102} {\bibfield  {journal} {\bibinfo
  {journal} {Phys. Rev. Lett.}\ }\textbf {\bibinfo {volume} {120}},\ \bibinfo
  {pages} {264102} (\bibinfo {year} {2018})}\BibitemShut {NoStop}%
\bibitem [{\citenamefont {Fruchart}\ \emph {et~al.}(2021)\citenamefont
  {Fruchart}, \citenamefont {Hanai}, \citenamefont {Littlewood},\ and\
  \citenamefont {Vitelli}}]{fruchart21}%
  \BibitemOpen
  \bibfield  {author} {\bibinfo {author} {\bibfnamefont {M.}~\bibnamefont
  {Fruchart}}, \bibinfo {author} {\bibfnamefont {R.}~\bibnamefont {Hanai}},
  \bibinfo {author} {\bibfnamefont {P.~B.}\ \bibnamefont {Littlewood}}, \ and\
  \bibinfo {author} {\bibfnamefont {V.}~\bibnamefont {Vitelli}},\ }\bibfield
  {title} {\enquote {\bibinfo {title} {Non-reciprocal phase transitions},}\
  }\href@noop {} {\bibfield  {journal} {\bibinfo  {journal} {Nature}\ }\textbf
  {\bibinfo {volume} {592}},\ \bibinfo {pages} {363--369} (\bibinfo {year}
  {2021})}\BibitemShut {NoStop}%
\bibitem [{\citenamefont {Montbri\'o}\ and\ \citenamefont
  {Paz\'o}(2018)}]{MP18}%
  \BibitemOpen
  \bibfield  {author} {\bibinfo {author} {\bibfnamefont {E.}~\bibnamefont
  {Montbri\'o}}\ and\ \bibinfo {author} {\bibfnamefont {D.}~\bibnamefont
  {Paz\'o}},\ }\bibfield  {title} {\enquote {\bibinfo {title} {Kuramoto model
  for excitation-inhibition-based oscillations},}\ }\href {\doibase
  10.1103/PhysRevLett.120.244101} {\bibfield  {journal} {\bibinfo  {journal}
  {Phys. Rev. Lett.}\ }\textbf {\bibinfo {volume} {120}},\ \bibinfo {pages}
  {244101} (\bibinfo {year} {2018})}\BibitemShut {NoStop}%
\bibitem [{\citenamefont {Laing}\ \emph {et~al.}(2021)\citenamefont {Laing},
  \citenamefont {Bl\"asche},\ and\ \citenamefont {Means}}]{laing21}%
  \BibitemOpen
  \bibfield  {author} {\bibinfo {author} {\bibfnamefont {C.~R.}\ \bibnamefont
  {Laing}}, \bibinfo {author} {\bibfnamefont {C.}~\bibnamefont {Bl\"asche}}, \
  and\ \bibinfo {author} {\bibfnamefont {S.}~\bibnamefont {Means}},\ }\bibfield
   {title} {\enquote {\bibinfo {title} {Dynamics of structured networks of
  {W}infree oscillators},}\ }\href {\doibase 10.3389/fnsys.2021.631377}
  {\bibfield  {journal} {\bibinfo  {journal} {Front. Syst. Neurosci.}\ }\textbf
  {\bibinfo {volume} {15}},\ \bibinfo {pages} {631377} (\bibinfo {year}
  {2021})}\BibitemShut {NoStop}%
\bibitem [{\citenamefont {Hong}\ and\ \citenamefont {Strogatz}(2011)}]{HS11}%
  \BibitemOpen
  \bibfield  {author} {\bibinfo {author} {\bibfnamefont {H.}~\bibnamefont
  {Hong}}\ and\ \bibinfo {author} {\bibfnamefont {S.~H.}\ \bibnamefont
  {Strogatz}},\ }\bibfield  {title} {\enquote {\bibinfo {title} {Kuramoto model
  of coupled oscillators with positive and negative coupling parameters: An
  example of conformist and contrarian oscillators},}\ }\href {\doibase
  10.1103/PhysRevLett.106.054102} {\bibfield  {journal} {\bibinfo  {journal}
  {Phys. Rev. Lett.}\ }\textbf {\bibinfo {volume} {106}},\ \bibinfo {pages}
  {054102} (\bibinfo {year} {2011})}\BibitemShut {NoStop}%
\bibitem [{\citenamefont {Uchida}\ and\ \citenamefont
  {Golestanian}(2010)}]{uchida10}%
  \BibitemOpen
  \bibfield  {author} {\bibinfo {author} {\bibfnamefont {N.}~\bibnamefont
  {Uchida}}\ and\ \bibinfo {author} {\bibfnamefont {R.}~\bibnamefont
  {Golestanian}},\ }\bibfield  {title} {\enquote {\bibinfo {title}
  {Synchronization in a carpet of hydrodynamically coupled rotors with random
  intrinsic frequency},}\ }\href {\doibase 10.1209/0295-5075/89/50011}
  {\bibfield  {journal} {\bibinfo  {journal} {Europhys. Lett.}\ }\textbf
  {\bibinfo {volume} {89}},\ \bibinfo {pages} {50011} (\bibinfo {year}
  {2010})}\BibitemShut {NoStop}%
\bibitem [{\citenamefont {Hanai}(2022)}]{hanai22}%
  \BibitemOpen
  \bibfield  {author} {\bibinfo {author} {\bibfnamefont {R.}~\bibnamefont
  {Hanai}},\ }\bibfield  {title} {\enquote {\bibinfo {title} {Non-reciprocal
  frustration: time crystalline order-by-disorder phenomenon and a
  spin-glass-like state},}\ }\href@noop {} {\bibfield  {journal} {\bibinfo
  {journal} {arXiv preprint arXiv:2208.08577v2}\ } (\bibinfo {year}
  {2022})}\BibitemShut {NoStop}%
\bibitem [{\citenamefont {Mart\'{\i}}\ \emph {et~al.}(2018)\citenamefont
  {Mart\'{\i}}, \citenamefont {Brunel},\ and\ \citenamefont
  {Ostojic}}]{dani18}%
  \BibitemOpen
  \bibfield  {author} {\bibinfo {author} {\bibfnamefont {D.}~\bibnamefont
  {Mart\'{\i}}}, \bibinfo {author} {\bibfnamefont {N.}~\bibnamefont {Brunel}},
  \ and\ \bibinfo {author} {\bibfnamefont {S.}~\bibnamefont {Ostojic}},\
  }\bibfield  {title} {\enquote {\bibinfo {title} {Correlations between
  synapses in pairs of neurons slow down dynamics in randomly connected neural
  networks},}\ }\href {\doibase 10.1103/PhysRevE.97.062314} {\bibfield
  {journal} {\bibinfo  {journal} {Phys. Rev. E}\ }\textbf {\bibinfo {volume}
  {97}},\ \bibinfo {pages} {062314} (\bibinfo {year} {2018})}\BibitemShut
  {NoStop}%
\bibitem [{\citenamefont {Berlemont}\ and\ \citenamefont
  {Mongillo}(2022)}]{mongillo}%
  \BibitemOpen
  \bibfield  {author} {\bibinfo {author} {\bibfnamefont {K.}~\bibnamefont
  {Berlemont}}\ and\ \bibinfo {author} {\bibfnamefont {G.}~\bibnamefont
  {Mongillo}},\ }\bibfield  {title} {\enquote {\bibinfo {title} {Glassy phase
  in dynamically-balanced neuronal networks},}\ }\href {\doibase
  10.1101/2022.03.14.484348} {\bibfield  {journal} {\bibinfo  {journal}
  {bioRxiv}\ } (\bibinfo {year} {2022}),\
  10.1101/2022.03.14.484348}\BibitemShut {NoStop}%
\bibitem [{Note1()}]{Note1}%
  \BibitemOpen
  \bibinfo {note} {In that case $\protect \mathrm {rank}({\protect \mathbf
  M})=L$ since $M_{jk}=\protect \frac 12({\protect \bf u}_j - {\protect \bf
  v}_j)\cdot ({\protect \bf u}_k + {\protect \bf v}_k)$.}\BibitemShut {Stop}%
\bibitem [{\citenamefont {Ott}\ and\ \citenamefont {Antonsen}(2008)}]{OA08}%
  \BibitemOpen
  \bibfield  {author} {\bibinfo {author} {\bibfnamefont {E.}~\bibnamefont
  {Ott}}\ and\ \bibinfo {author} {\bibfnamefont {T.~M.}\ \bibnamefont
  {Antonsen}},\ }\bibfield  {title} {\enquote {\bibinfo {title} {Low
  dimensional behavior of large systems of globally coupled oscillators},}\
  }\href {\doibase 10.1063/1.2930766} {\bibfield  {journal} {\bibinfo
  {journal} {Chaos}\ }\textbf {\bibinfo {volume} {18}},\ \bibinfo {eid}
  {037113} (\bibinfo {year} {2008})}\BibitemShut {NoStop}%
\bibitem [{Note2()}]{Note2}%
  \BibitemOpen
  \bibinfo {note} {For completeness: The imaginary eigenvalues of ${\protect
  \bm {\protect \mathcal {M}}}$ are $\pm i (1-\eta ) 2^{4L-1}$. A sketch of the
  proof follows. We denote the normalized eigenvectors of $\protect \bm
  {\protect \mathcal {S}}$ with positive (negative) eigenvalue by ${\protect
  \mathbf s}_i^+$ (${\protect \mathbf s}_i^-$) . From the identities ${\protect
  \bm {\protect \mathcal {S}}}^2+{\protect \bm {\protect \mathcal
  {A}}}^2={\protect \bm {\protect \mathcal {SA}}}+{\protect \bm {\protect
  \mathcal {AS}}}=0$, we find that ${\protect \bm {\protect \mathcal {A}}}
  {\protect \mathbf s}_i^{+}=2^{2L} {\protect \mathbf s}_i^{-}$ and ${\protect
  \bm {\protect \mathcal {A}}} {\protect \mathbf s}_i^{-}=- 2^{2L} {\protect
  \mathbf s}_i^{+}$. The eigenvectors of ${\protect \bm {\protect \mathcal
  {A}}}$ are of the form ${\protect \mathbf s}_i^{+}\pm i{\protect \mathbf
  s}_i^{-}$, and the associated eigenvalues are $\pm i 2^{2L}$, completing in
  this way the proof.}\BibitemShut {Stop}%
\bibitem [{Note3()}]{Note3}%
  \BibitemOpen
  \bibinfo {note} {See the Supplemental Material at [...] for
  details}\BibitemShut {NoStop}%
\bibitem [{\citenamefont {Mastrogiuseppe}\ and\ \citenamefont
  {Ostojic}(2018)}]{mastrogiuseppe}%
  \BibitemOpen
  \bibfield  {author} {\bibinfo {author} {\bibfnamefont {F.}~\bibnamefont
  {Mastrogiuseppe}}\ and\ \bibinfo {author} {\bibfnamefont {S.}~\bibnamefont
  {Ostojic}},\ }\bibfield  {title} {\enquote {\bibinfo {title} {Linking
  connectivity, dynamics, and computations in low-rank recurrent neural
  networks},}\ }\href {\doibase https://doi.org/10.1016/j.neuron.2018.07.003}
  {\bibfield  {journal} {\bibinfo  {journal} {Neuron}\ }\textbf {\bibinfo
  {volume} {99}},\ \bibinfo {pages} {609--623} (\bibinfo {year}
  {2018})}\BibitemShut {NoStop}%
\bibitem [{\citenamefont {Schuessler}\ \emph {et~al.}(2020)\citenamefont
  {Schuessler}, \citenamefont {Dubreuil}, \citenamefont {Mastrogiuseppe},
  \citenamefont {Ostojic},\ and\ \citenamefont {Barak}}]{schuessler}%
  \BibitemOpen
  \bibfield  {author} {\bibinfo {author} {\bibfnamefont {F.}~\bibnamefont
  {Schuessler}}, \bibinfo {author} {\bibfnamefont {A.}~\bibnamefont
  {Dubreuil}}, \bibinfo {author} {\bibfnamefont {F.}~\bibnamefont
  {Mastrogiuseppe}}, \bibinfo {author} {\bibfnamefont {S.}~\bibnamefont
  {Ostojic}}, \ and\ \bibinfo {author} {\bibfnamefont {O.}~\bibnamefont
  {Barak}},\ }\bibfield  {title} {\enquote {\bibinfo {title} {Dynamics of
  random recurrent networks with correlated low-rank structure},}\ }\href
  {\doibase 10.1103/PhysRevResearch.2.013111} {\bibfield  {journal} {\bibinfo
  {journal} {Phys. Rev. Res.}\ }\textbf {\bibinfo {volume} {2}},\ \bibinfo
  {pages} {013111} (\bibinfo {year} {2020})}\BibitemShut {NoStop}%
\end{thebibliography}

%

\clearpage

\begin{widetext}
\begin{center}
 {\large\bf{Supplemental Material to}}\\

{\large\bf{Volcano transition in populations of phase oscillators with random nonreciprocal interactions}}
\end{center}

We have redone the simulation by Kimoto and Uezu (K\&U) \cite{kimoto19}, using the same population size $N=500$, and a coupling strength
$\tilde J=1$. Following \cite{kimoto19} the frequency dispersion $\sigma$ was decreased quasiadiabatically,
with a step size $\Delta\sigma=1.5\times 10^{-4}$, slightly smaller than $\Delta\sigma=\sqrt{\pi/2}\times 1.25\times 10^{-4}
\simeq1.567\times 10^{-4}$ used in \cite{kimoto19}. The initial $\sigma$ value was 1.88, almost identical
to $\sqrt{\pi/2}\times 1.5\simeq1.87997$ used in \cite{kimoto19}. Computation time for each $\sigma$ value
was 800 t.u. long, while 10000 t.u. (recording the local fields every time unit) were run at specific $\sigma$ values, as in \cite{kimoto19}.

The results for two completely independent numerical simulations are shown in Fig.~\ref{fig}.
One data set (crosses) are the locations of the maximum of the histogram of local fields amplitudes,
while the red circles are the values of $r_*$ after fitting the histogram to Eq.~(27) in the main text.
The prediction by K\&U (coincident with our extrapolation from a low-rank coupling matrix)
is
\begin{equation}
1=\frac{2}{\pi g(0)}=   \sqrt{\frac{8}\pi} \sigma_v \qquad   \Rightarrow \qquad \sigma_v= \sqrt{\frac\pi8}\simeq 0.627 ,
\end{equation}
see dashed line in Fig.~\ref{fig}. In our view there is a nonnegligible discrepancy between theory and numerics.
Alternatively, we can infer $\sigma_v$ from our numerical result in Fig.~5 with $N=400$.
There we fixed $\sigma=\sqrt{\pi/2}$ and varied  $\tilde J$. Now, we can move to $\sigma$ space obtaining
\begin{equation}
\sigma_v^{emp}=\frac{\sqrt{\pi/2}}{{\tilde J}_v^{emp}},
\label{sv}
\end{equation}
where ${{\tilde J}_v^{emp}}$ denotes the empirical critical coupling for the volcano transition in Fig.~5.
According to Fig. 5 in the main text criticality is in the range $1.6<{\tilde J}_v^{emp}<1.7$, so we expect
\begin{equation}
0.737<\sigma_v^{emp}< 0.783
\end{equation}
This estimation is compatible with the results in Fig.~\ref{fig}.
Nonetheless, an extensive study with more simulations, and ideally with larger systems sizes is probably in order.

\begin{figure}[h]
\begin{minipage}[t]{0.40\linewidth}
\includegraphics[width=\textwidth]{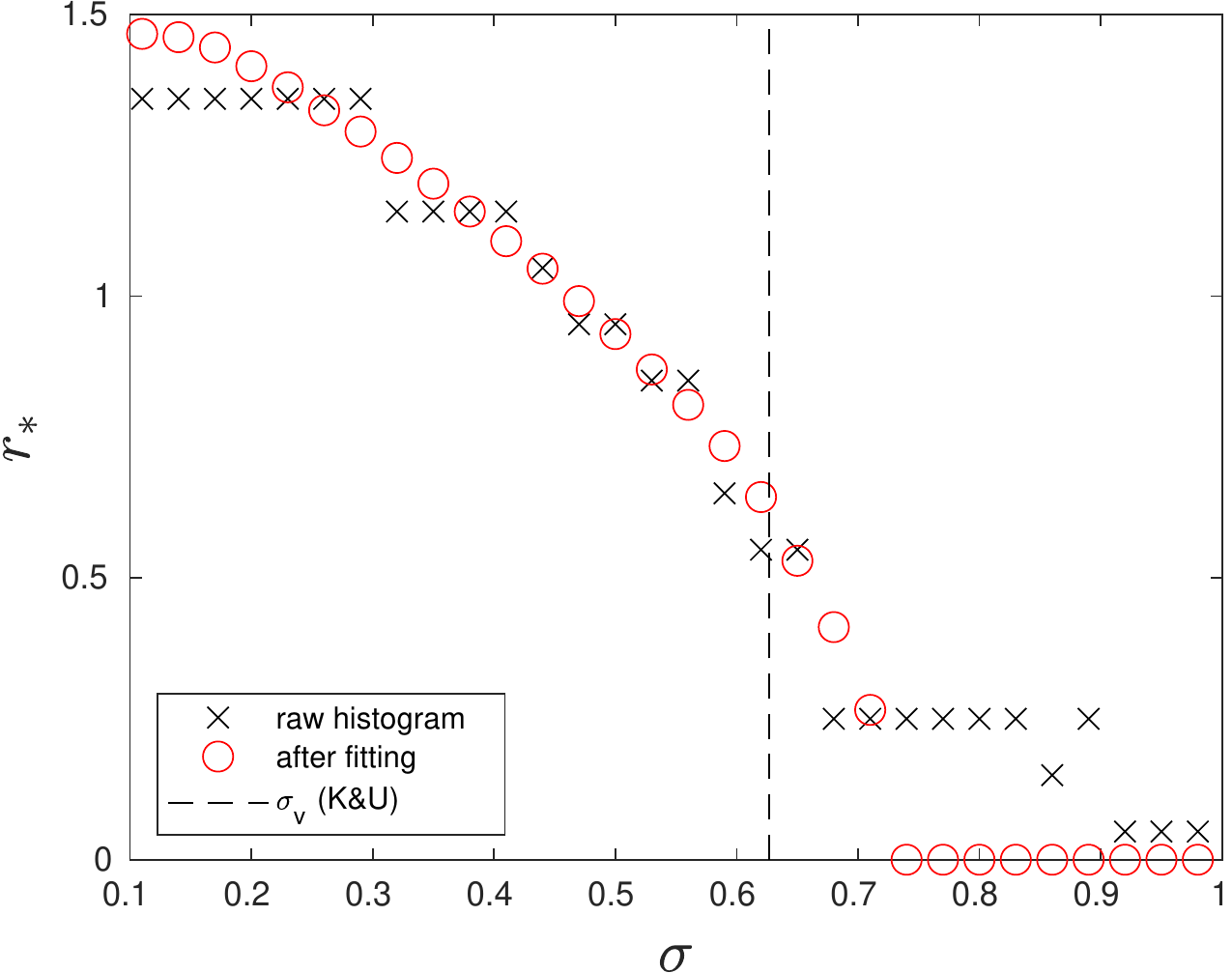}
\end{minipage}
\begin{minipage}[t]{0.45\linewidth}
\includegraphics[width=\textwidth]{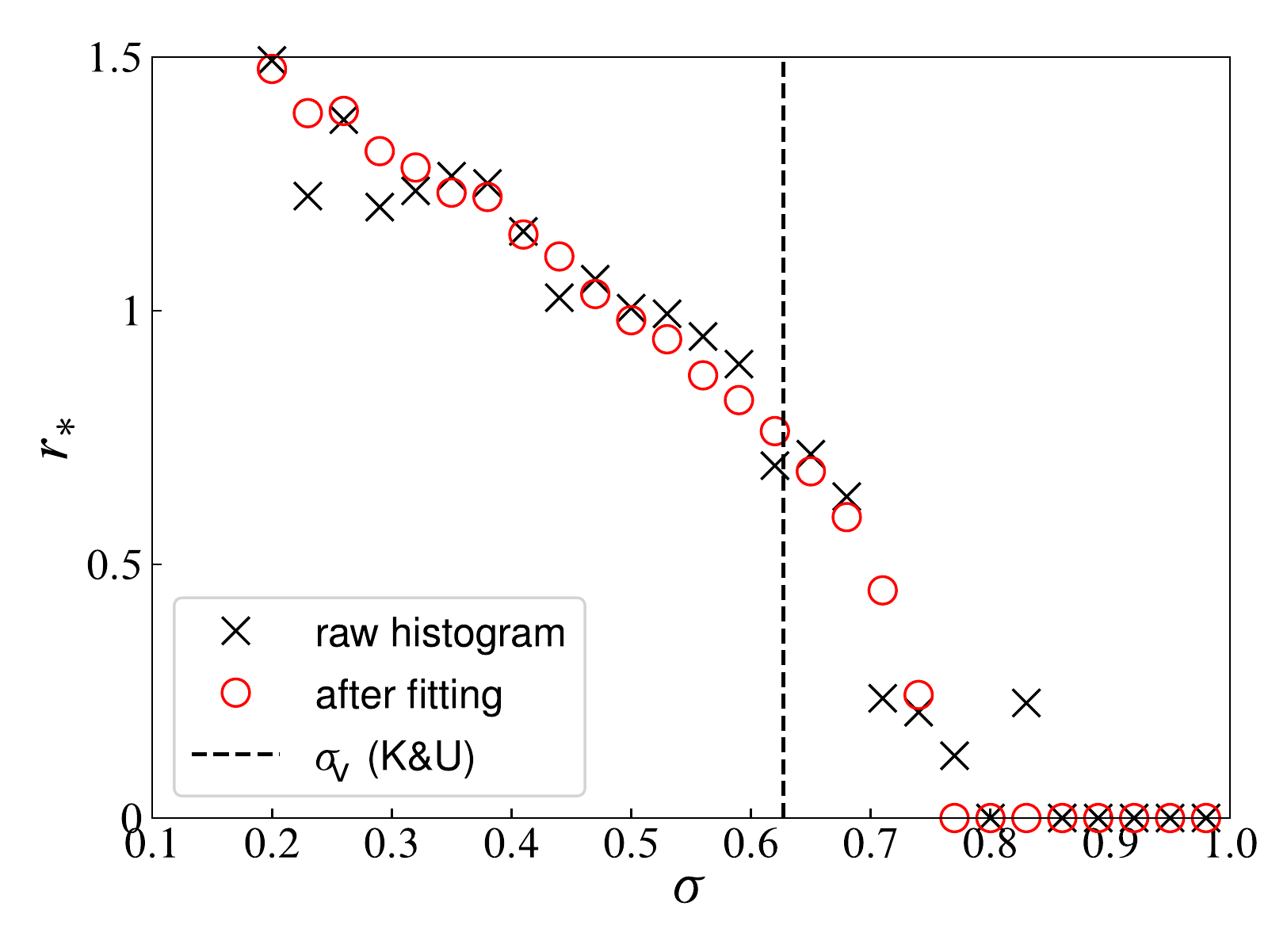}
\end{minipage}
\caption{Maximum of the distribution of local fields $r_*$ as a function of the frequency dispersion.
Each panel corresponds to a completely independent numerical implementation (one by each author)
using a 4th order Runge-Kutta algorithm of step size 0.1.
Confront with Fig.~6(b) of \cite{kimoto19}.}
\label{fig}
\end{figure}

\end{widetext}

\end{document}